\documentstyle[11pt,newpasp,twoside,psfig]{article}
\def\edcomment#1{\iffalse\marginpar{\raggedright\sl#1\/}\else\relax\fi}
\reversemarginpar

\begin{document}
\title{Four Partially Baked Results}
 \author{Robert Antonucci}
\affil{University of California, Santa Barbara, Department of Physics, Santa Barbara, CA  93106}

\author{David H. Whysong}
\affil{University of California, Santa Barbara, Department of Physics, Santa Barbara, CA  93106}

\setcounter{page}{111}
\index{Robert Antonucci}
\index{David Whysong}

\begin{abstract}
D Whysong and I have been carrying out a mid-IR survey of 3C radio galaxies to
see which have IR-excesses, interpretable as hidden Type-1 AGN.  Mostly due to
relentless bad weather, we have a modest collections of observations so far.
However, we can say with confidence that some radio galaxies are \textit{true
kinetic, nonthermal emitters}, with negligible energetic contribution from a
visible or hidden Big Blue Bump

To make up for the modest collection of mid-IR observations so far, we also
present a new spectropolarimetric discovery regarding the underlying shape of
the quasar Big Blue Bump in the Balmer Edge region, a fairly radical view of
quasar reddening, and a bit of Keck adaptive optics data on Cygnus A and the
ultraluminous infrared galaxy NGC6240.
\end{abstract}

\section{Introduction}

My topic was to be Thermal Emission as a Test for Hidden \textit{AGN} in Radio
Galaxies, work in collaboration with, and in fact largely carried out by UCSB
graduate student Dave Whysong.  As I will explain, we are attempting to
determine which FRII radio galaxies in the 3C catalog are likely to have hidden
quasars by the consequent reradiation of energy by warm dust.  The answer is
far from all, and the dependence on optical spectrum, radio size, etc is
essential for the Unified Model, and for black hole accretion mechanisms and
demographic studies.  Since our data set is insufficiently complete, I will
emphasize instead some nearby individual objects of interest.  These latter are
discussed in astro-ph/0207385, a manuscript submitted to ApJ.

Four partially baked (about half actually) results together must suffice for
one fully baked one.  Therefore I've added some information on a recent
discovery by Makoto Kishimoto, Omer Blaes and myself, which reveals the
spectral shape of the Big Blue Bump for the first time in the important region
of the Balmer edge.  Next I show some results on the quasar reddening curve,
which are startling, and very important, or else wrong.  The authors there are
Martin Gaskell, Rene Goosmann, David Whysong, and myself.  Finally I preview
some adaptive optics work from Keck on Cygnus A and \textit{NGC6240}, courtesy
of collaborators Claire Max, Gabriella Canalizo, David Whysong, Bruce
Macintosh, and Alan Stockton.

This paper is obviously not a review but just a compact story that cites a
negligible fraction of the relevant literature.

\section{Thermal Emission as a Test for Hidden Quasars in Radio Galaxy Nuclei}

Supermassive black holes, or something that behaves a lot like them, lie at the
heart of radio galaxies, quasars, and other \textit{AGN}.  However, they manifest
themselves in a (formerly) bewildering variety of ways.  In particular, quasars
are powerful optical/UV emitters, and this emission is thought to arise somehow
in matter radiating as it approaches the hole's horizon.  As far as I know this 
''Big Blue Bump" relatively quiescent and low polarization continuum component
is always accompanied by commensurate broad emission lines.  Powerful radio
galaxies and radio quasars also have jets and lobes. It's been known or
suspected for 20 years that the most powerful radio galaxies actually contain
quasars hidden from our direct view, hidden by something that acts like a
geometrically thick torus oriented perpendicular to the radio jets.

My strong impression is that this is almost universally true at the very top
end of the radio galaxy luminosity function, where giant bipolar reflection
nebulae are seen whose polarized spectra show the quasar-like broad emission
lines.  Figures 1 and 2 in Singal 1993 (compare Barthel 1989) show that the
$Z>0.5$ radio galaxies in the culled 3C catalog bear the expected relationship to
the quasars in space density and projected linear size of the radio sources for
this simple picture.  (The ``culled" catalog is pretty cleanly those with enough
\textit{isotropic} flux to satisfy the minimum catalog flux requirement.)  At low
Z (which may simply mean lower luminosity) the simplest and most charitable
thing that can be said is that perhaps there are many radio galaxies of a new
type present, which do not have hidden quasars, and which spoil the statistical
relationships of the quasars to the radio galaxies.  This speculation was made
by many of us early on.

However a paper appeared soon after Singal's which taught me that I have no
critical faculty; that all the statistics are just as expected if more powerful
nuclei tend to have larger torus opening angles, and if in addition radio
sources tend to fade over time (Gopal-Krishna et al 1996).  In Singal's low
redshift/luminosity bin our eyes are then comparing old quasars to younger
radio galaxies.  Clearly this idea can't be tested against the idea of a new
population of small ``true" radio galaxies by reason alone.  Spectropolarimetry
affords a robust detection of a hidden quasar in favorable cases, but if the
opening angle is small or zero, or if insufficient scattering material is
visible, it simply returns a false negative answer.  But if the initial Big
Blue Bump radiation is even roughly isotropic, and if the reradiation by the
torus is even roughly isotropic (there are good arguments for both of these),
then the ``waste heat" from the hidden nucleus must appear as thermal emission
from warm dust.  Experience and theory show that the mid-IR provides the best
combination of isotropy and contrast relative to any cooler dust heated by a
starburst component.

Our initial results on special targets were presented in a conference in 2000,
and made public in Mar 2001 as astro-ph/0103048 (Antonucci 2002); Whysong and
Antonucci 2001 (astro-ph/0106381, June 2001);  and Whysong and Antonucci 2002
(astro-ph/0207385, sub to ApJ).  I'm specific here because of some puzzling
subsequent statements by other authors.  Perlman et al (2001) discussed our
result showing the lack of an energetically significant hidden Big Blue Bump in
M87, refined it slightly, and then with Gemini Observatory issued a press
release strongly suggesting it was theirs.\footnote{The Perlman et al. data have
somewhat lower spatial resolution but much better surface brightness
sensitivity to off nuclear emission when compared with ours.  The authors argue
that the latter gives them better limits on nuclear reradiated emission, but at
these luminosity levels the reradiated mid-IR would be spatially unresolved.}
Radomski et al (2002) use our Cygnus A photometry in their paper, but publish
and analyze their image without reference to our similar image and brief
analysis in the paper of ours which they cite.

Very briefly our conclusions on the nearby objects are 1) M87 shows a point
source of 13 mJy at 11.7 microns, consistent with synchrotron emission from the
radio core.  Any reradiated thermal mid-infrared emission is several orders of
magnitude weaker than the kinetic power.  Arguably this is the first near-proof
of the existence of a ``nonthermal \textit{AGN},"  i.e. an \textit{AGN} with a
negligible fraction of its total power output as optical/UV emission.  Cygnus A
on the other hand has a fairly powerful and extended infrared source, as
expected for this object in which a hidden quasar has been detected via
spectropolarimetry (Ogle et al 1997).

We found that Cen A has a point source at wavelengths 11.7 and 17.7 microns for
which the slope and spectrum is suggestive of thermal reradiation from a hidden
Big Blue Bump. In this case there is supportive but not definitive polarimetric
evidence for a hidden low-luminosity quasar (Marconi et al 2000).  Also our
point source flux at ~0.3" resolution flux matches that measured by
\textit{ISO} in a 4" aperture; the \textit{ISO} spectrum, which thus is also a
spectrum of our point source, has strong dust absorption features, and pretty
convincing dust emission features as well (Mirabel et al 1999).  Comparing M87
to Cen A, a key conclusion is that some FR I radio galaxies have the hidden Big
Blue Bump and some do not!\footnote{A pet peeve of mine is the constant refrain
in the literature stating that BL Lacs are beamed FR1 radio sources;  as has
been known for a decade or two, depending on how you count it, this is not true
of the ones which have FR2 diffuse radio emission.  Similarly I read now that
FR1s are generically nonthermal (no hidden Big Blue Bump).  This is in fact
true for all of them except the ones that are thermal and do have a visible or
hidden \textit{BBB}.}

\section{All Quasars are Blue}

I have undergone a religious conversion regarding quasar reddening.  With the
zeal of a convert, I believe that most quasars are reddened in the optical, and
absorbed but not reddened in the UV.  Despite my zeal, I still have some
doubts, but they can be addressed robustly in the near future.

As usual there are many prescient papers on this subject\footnote{Early
examples include Osterbrock, Koski and Phillips 1976 for radio galaxies' broad
line regions and continua, and de Zotti and Gaskell 1985 for dust in the host
galaxy planes of Seyferts.  Some especially telling later work on radio
galaxies includes Hill, Goodrich and dePoy 1996 and Hines et al 1999, but again
these comprise a small fraction of the literature.}, but I'll again feature
just a few landmarks from my personal reading odyssey.  J Baker and colleagues
wrote a series of papers a few years back on the radio and optical properties
of the Molonglo radio survey quasars, a sample selected at low frequencies and
with very good completeness and follow-up of the optical identifications.  Most
relevant here are Baker and Hunstead 1995, Baker 1997, and Baker et al 1999.

For the 13 objects with broad line Balmer decrements available, the quasars
showed a good correlation between that parameter and the optical slope (Baker
1997, Fig 16).  The authors interpreted this indicative of reddening.  The
consequences were profound and perhaps not widely appreciated.  Such a general
effect should have major implications for modeling of the Big Blue Bump, the
energetically dominant continuum component.  In particular, it implies that
almost all these radio loud quasars are intrinsically quite blue in the
optical, and theorists should feel no obligation to model the steeper ones.  On
the other hand, it means that observed trends in optical slope, e.g.
correlations with luminosity, can not be interpreted with confidence as effects
of changing luminosity or Eddington ratio.

The impact is equally profound for quasar and black hole demographic studies:
such studies universally assume that broad line object luminosities are just
about what we observe them to be.  If the optical slopes generally have a
non-negligible signature of reddening, then even with the reddening law we
discuss below, \textit{Most of the optical/UV luminosity is absorbed by dust}. Thus
both their luminosities and the luminosities of the hidden ones must be revised
upward substantially.

I suggested in Antonucci 2002 (most easily accessed as astro-ph/0103048) that
the optical slopes could not in fact be so influenced by reddening without
seeing the sharp characteristic curvature of foreground reddening at the
short
 wavelength end of the published spectra\footnote{It's common for
intermixed stars and gas to show optical reddening without an exponential UV
cutoff, but it's quite novel for the case of a point source and
\textit{foreground absorption}}. I thought using the single line pair broad
H-alpha/broad H-beta was risky, and that that particular pair might be
influenced by ionization parameter effect.  After discussing this in detail
with Martin Gaskell and Rene Goosman, those workers showed me that \textit{all}
broad line ratios agree with the reddening notion, and that it's possible to
derive nearly identical continuum and emission line reddening curves using all
six well measured emission lines.  Physically, this simply means that the small
grains are efficiently destroyed; there are many independent arguments for
that.  Our ``universal" (radio loud) quasar reddening curve is shown in Fig 1 of
Gaskell et al 2002.

The beauty of all this is that absorption of this magnitude can be readily
recognized by the concommitent infrared luminosity, which almost
necessarily accompanies it.  We are gathering data on this now.

\section{Se Raser la Barbe}

A completely frustrating aspect of the quasar spectrum is that the underlying
continuum with its diagnostic features is very heavily contaminated by
overlying atomic emission from the broad emission line region.  There is a way
to get rid of it however!  At least a few quasars have a small
wavelength-independent polarization in the bits which are thought to be
uncontaminated by atomic emission.  At the same time, these objects have
undetectably low polarization of the broad emission lines.  Because of the
latter feature, a plot of polarized flux acts to shave off all the atomic
emission as one can shave off a beard with a razor blade.  Note that the cause
of the polarization needn't be known;  I think of it as the stinky stuff they
add to natural gas to make it noticeable.  It's worth noting however that the
polarization angle is parallel to the radio structure axis, suggestive of light
both generated and scattered within a thin disk; it is not consistent with
light scattered in a photosphere above a glowing thin disk (Stockman, Angel and
Miley 1979).

Of particular interest is the Balmer edge region.  Kolykhalov and Sunyaev
(1984) on the theoretical side, and our group on the observational side
(Antonucci, Kinney and Ford 1989), started a cottage industry trying to use the
Ly edge feature (or lack of it) to diagnose the Big Blue Bump emission
mechanism.  However, in a thermal model such as a thin accretion disk, the Ly
edge region can be substantially affected by relativisitic broadening and
shifting effects;  it's also sensitive to the non-LTE conditions thought to
prevail in the innermost regions;  and it depends on the physics in a region
subject to various classical instabilities.  All these problems are much
mitigated for the case of the Ba edge.  It hasn't been possible to study the
Big Blue Bump Balmer edge region observationally however, because of an enormous
relatively localized atomic feature called the Small Blue Bump.  But we can
shave that off with the polarized flux method.

Several quasars are known to have continuum polarization, but little or no
polarization in their broad emission lines and Small Blue Bump features (data
of Miller and Goodrich in Antonucci 1988;  Antonucci et al 1996; Schmidt and
Smith 2000).  Using the Keck and VLT observatories we can now get more detailed
data and our fearless leader M. Kishimoto (these proceedings)
shows that the quasar Ton 202 has an unpolarized Small Blue Bump.  More
importantly, the data reveal that the underlying Big Blue Bump continuum
component has a break at the exact same wavelength that the Small Blue Bump
arises, namely around 4000A in the rest frame!  This probably means that the
former is thermal and optically thick.

A consequence is that the Small Blue Bump, which is already quite
problematically huge to explain with the usual Balmer line + Fe II
interpretation, is even huger since it lies in a region deficient in underlying
continuum flux.  This last conclusion would be affected though if the deficit
in polarized flux derives from localized diminished percent polarization in the
Big Bump rather than diminished flux.  Theoretical discussions of the
wavelength-dependence of accretion disk polarization can be found in for
example Laor, Netzer and Piran 1990, with the state of the art being Agol,
Blaes and Ionescu-Zanetti 1998.

Note that the data do not indicate any particular physics for the Big Bump
however;  at this point we can only say they probably indicate an origin in
optically thick thermal matter.  Of course this new information does not
negate the many really damning arguments against the Shakura-Sunyaev
disk and some of its simple variants (e.g. Antonucci 1999).

\section{V  Better than Hubble? Keck Adaptive Optics Observations of Cygnus A and
NGC6240}

This section refers to work by Claire Max - the captain of this enterprise, G
Canalizo, D Whysong, B Macintosh, and myself, and by Claire with other
collaborators (see author lists).

Is a big ground-based telescope sharper for imaging and spectroscopy than the
Hubble?  The answer is of course yes if the same wavelength (and A.O.) is used;
however, the dynamic range and the field size are both lower.  Fig 1 shows the
core region of Cygnus A imaged at 2 microns by HST and by Keck, using a
fortuitously placed natural guide star.  Clearly the Keck data are at
considerably higher resolution as well as taken with good sampling, yet our
restoration is still necessarily very conservative relative to the diffraction
limit.  There is a published Abstract on this: Max et al 2001; and much more
and sharper data are in hand!

To my biased eye, this and some published images (for example Jackson et al
1998, Tadhunter et al 1999, Fosbury et al 1999) look like you can reach right
out and touch Cygnus, and in particular it looks like various parts of the
Humunculus nebula associated with eta Carinae.  Our data also include
high-resolution spectra, but I am truly running out of space, so I'll just
close with a mention of the spectrum of NGC6240.

This is another prototypical object by which the good gods have placed a
natural guide star.  NGC6240 is a prototype ULIRG double galaxy with powerful
AGN and starburst activity. Lacking room to show the images and slit spectra,
I'll just have to note that they show the $H_2$ thermal line, and also Fe II
characteristic of Liners in exquisite detail.  A little more information is
available in Max et al 2000 and Bogdanovic et al 2001;  a paper is in
preparation.

\bigskip

\begin{figure}[t]
\centerline{
\psfig{figure=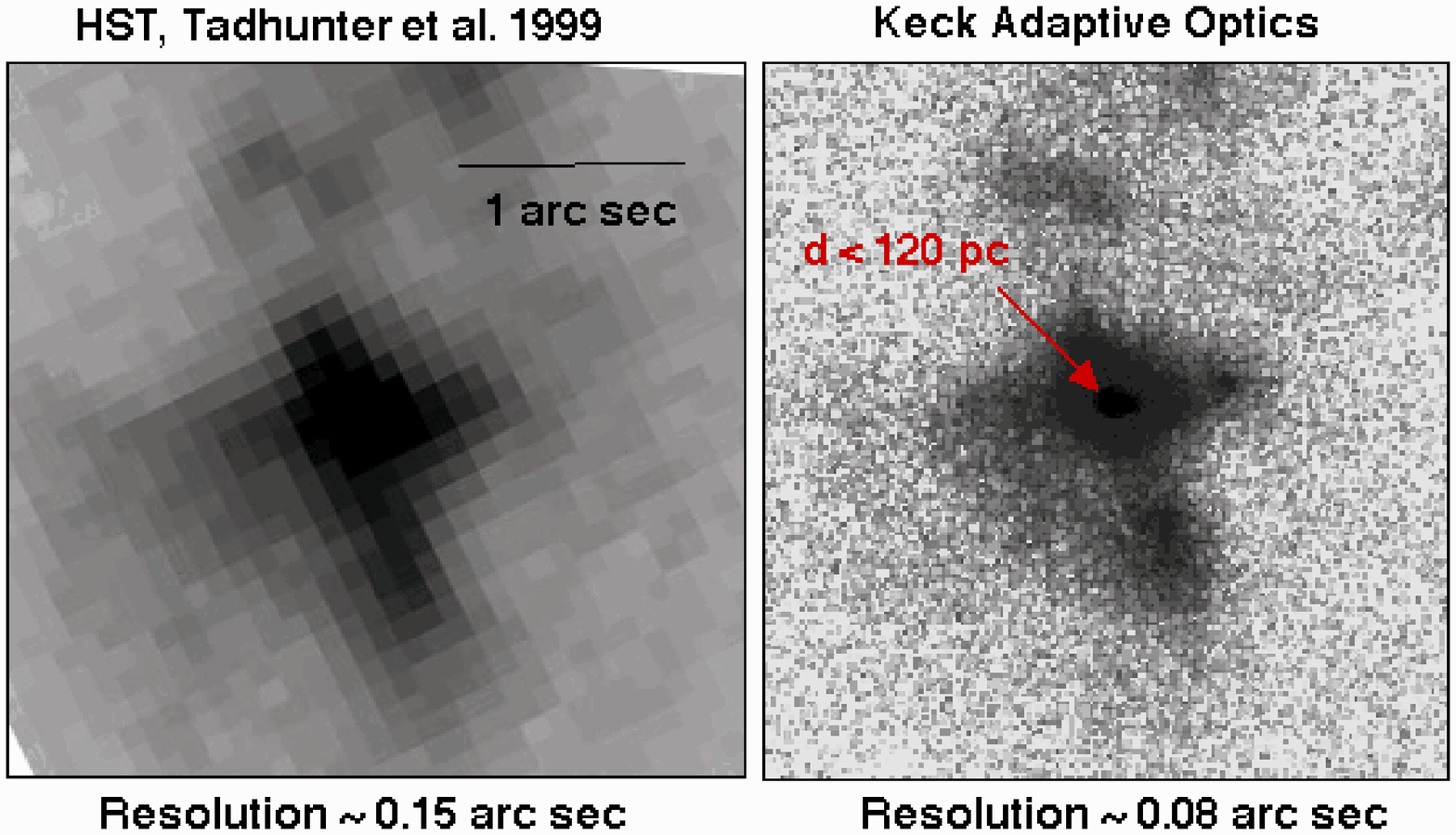,width=11cm,angle=-90}
}
\caption{ }
\label{fig1}
\end{figure}

\vspace{.20in}
\acknowledgements
Thanks are due to R Barvainis for commenting on an earlier verion of this
paper.

\end{document}